\documentstyle[preprint,eqsecnum,aps]{revtex}
\input epsf
\begin{document}
\preprint{MC/TH 96/07}
\title{Baryon masses in a chiral expansion with meson-baryon\\
form factors}
\author{Richard E. Stuckey and Michael C. Birse}
\address{Theoretical Physics Group, Department of Physics and Astronomy\\
University of Manchester, M13 9PL, UK\\}
\maketitle

\begin{abstract}
The chiral expansion of the one-loop corrections to baryon masses is examined
in a generic meson-cloud model with meson-baryon form factors. For pion loops,
the expansion is rapidly convergent and at fourth order in $m_\pi$ accurately
reproduces the full integral. In contrast, the expansion is found to converge
very slowly for kaon loops, raising questions about the usefulness of chiral
expansions for kaon-baryon physics. Despite the importance of high-order
terms, relations like that of Gell-Mann and Okubo are well satisfied by the
baryon masses calculated with the full integral. The pion cloud cloud makes a
significant contribution to the $\pi N$ sigma commutator, while kaon cloud
gives a very small strangeness content in the nucleon.
\end{abstract}

\section{Introduction}

The cloud of virtual mesons that surrounds any baryon contributes to the mass
and other properties of that particle. This can significantly change these 
properties compared to expectations based on simple quark models in which the
baryon is described as a three-quark state. Of particular importance in this
context are the pseudoscalar mesons, pions and kaons, since they form the
longest-ranged components of that cloud. 

These lightest mesons are approximate Goldstone bosons, whose masses arise
from the breaking of chiral symmetry by the current masses of the quarks.
Their contributions can thus be systematically expanded in powers of the
current-quark masses using the techniques of chiral perturbation theory (ChPT)
\cite{wei79,gl82,gl84,dgh,meirev,bkmrev}. Although this has been applied to
meson-baryon interactions with Dirac nucleons\cite{gss88}, the complications
introduced by the finite baryon mass mean that most practical applications
are based on heavy-fermion effective field theory\cite{ehg}. In this the
baryons are treated as very heavy and the effective action is expanded in
powers of the baryon momenta around the nonrelativistic limit\cite{jm91a}. All
ChPT approaches are based on nonrenormalisable effective Lagrangians; hence,
as the expansion is taken to higher orders, more and more counterterms are
introduced with coefficients that need to be determined.

The lowest-order approximations to heavy-baryon ChPT have much in common with
the cloudy bag model\cite{ttm80,thorev}: static or slow-moving baryons
perturbatively dressed with a meson cloud. The cloudy bag model and similar
chiral versions of the colour-dielectric soliton
model\cite{wd88,mcg91,dd93,bmcg96} are able to give very good descriptions of
low-energy baryon properties. Even if they cannot be extended to a complete
and rigorous chiral expansion, these models can also provide useful estimates
of the coefficients of some of the higher-order counterterms that will be
present in such an expansion. In particular these models give rise to form
factors at the meson-baryon vertices that regulate the loop integrals involved
in calculating the mesonic dressing of baryons.

Within the framework of a cloudy-bag approach, we have examined the chiral
expansion of the self-energies of octet and decuplet baryons including
one-loop contributions from the octet of pseudoscalar mesons (pions, kaons and
$\eta$). To look for general features of the approach we have used a generic
form factor instead of one calculated from the quark wave functions of some
specific bag or soliton model. We find that, for reasonable choices of form 
factor, the chiral expansion of pion loops converges rapidly. The results 
indicate that inclusion of terms up to fourth order in $m_\pi$ are sufficient 
to accurately reproduce the full integral. However the kaon loops converge 
much more slowly and terms up to at least seventh order in $m_K$ being needed.
The same pattern of an alternating series with very slow convergence is also
found in the work of Borasoy and Meissner\cite{bm96}, who calculated the
nucleon mass up to fourth order in full SU(3) ChPT. Our results indicate that
such calculations are likely to have to be extended to at least seventh order,
raising serious doubts about their feasibility. In contrast ChPT calculations
with nucleons (and $\Delta$'s, see below) and virtual pions are likely to be
well converged at the order currently achieved.

As in the cloudy-bag model\cite{thorev}, we have explicitly included the
decuplet of spin-${3\over 2}$ baryons in our calculations. These can play a
significant role in the mesonic dressing of the octet baryons because the
octet-decuplet splitting is comparable in magnitude to the pion mass. Indeed
Jenkins and Manohar\cite{jm91b} have argued that decuplet baryon fields
should be included in ChPT. This has been applied to calculations of baryon 
masses\cite{jen92,ll94} and $\sigma$-commutators\cite{jm92}. However as has
been pointed in Refs.\cite{bkm93,bm94} inclusion of the decuplet leads to new,
unknown counterterms at fourth order in ChPT.

The baryon decuplet plays a particularly important role in the limit where the
number of colours $N_c\rightarrow\infty$. It can then be thought of as the
first excited level of a rotational band based on a hedgehog intrinsic state,
as in the Skyrme\cite{sky,zbrev} and NJL soliton models\cite{njlrev}. In that
limit the octet-decuplet splitting vanishes as $1/N_c$ and so the chiral
expansion must be modified\cite{cb92}. The large-$N_c$ limit also leads to
various relations amongst baryon masses, which are independent of the details
of baryon structure\cite{djm94,jl95,jen95}.

Various baryon mass relations, normally regarded as tests of first-order
perturbation theory in the SU(3)-breaking\cite{gmo}, are found to be
very well satisfied by the full baryon-energies calculated with form factors.
Low-order terms in the chiral expansion would suggest much larger violations
of these relations, but these are not reliable because of the very slow
convergence of the expansion. In fact we find small violations of the GMO
relation, similar to those actually observed. Thus the success of the GMO
relation cannot be taken to imply the higher-order terms in the chiral
expansion are small. Another illustration of this is the very small
strange-quark content in the proton that we find in this approach.

Finally, the dressing of baryons with pseudoscalar mesons on its own does not
give a very good description of the full octet and decuplet spectrum, as noted
by McGovern\cite{mcg91}. We have therefore examined the effects of
symmetry-breaking terms and couplings beyond those of simple SU(6) quark-model
wave functions. Such effects can arise when, for example, gluon exchange 
forces are included in these models\cite{mcg91,dd93}. Our general conclusions
are unaffected by the inclusion of such terms.

\section{Chiral expansion of self-energies}

In a generic bag or soliton model, we take the bare mass of an octet or 
decuplet baryon to be of the form
\begin{equation}
M_A^{(0)}=M_0+N_s \epsilon_s+\delta,
\label{mbare}
\end{equation}
where $N_s$ is the number of strange quarks present in the baryon, $\epsilon_s$
the additional energy associated with a strange quark, $\delta$ is the bare
octet-decuplet mass splitting for a decuplet baryon and zero for an octet one.
(In the present work we do not consider isospin-breaking contributions to the
energy.) The mass of a dressed baryon can be written as
\begin{equation}
M_A=M_A^{(0)}+\Sigma_A,
\label{mdress}
\end{equation}
where $\Sigma_A$ is the self-energy of the baryon arising from its meson cloud.
At one-loop level in the cloudy-bag approach\cite{thorev} the self-energy of a
baryon A is a sum of terms of the form
\begin{equation}
\Sigma_{AB\nu} = -\frac{3}{4\pi^2}\left(\frac{g_{\pi NN}}{2M_N}\right)^2
\frac{(f^{AB\nu})^2}{25} \int_0^{\infty}\frac{k^4\mu^2(k)dk}
{\omega_\nu(k)\left(\omega_\nu(k)+M_B^{(0)}-M_A^{(0)}\right)}, 
\label{fullself}
\end{equation}
where $g_{\pi NN}$ is the pion nucleon coupling constant, $\mu(k)$ is the 
normalised meson-baryon form factor, and the energy of a meson of momentum $k$
and mass $m_\nu$ is
\begin{equation}
\omega_\nu(k)=\sqrt{k^2+m_\nu^2}.
\end{equation}
The factor $f^{AB\nu}$ is the coupling coefficient of baryon A to an
intermediate state consisting of baryon B and meson $\nu$. Values of these are
are listed in Table \ref{fab}, which generalises the similar table
in\cite{mmt86} by including kaon and $\eta$ couplings\cite{bae93} and allowing
for an $F/D$ ratio other than the 2/3 of an SU(6) quark model. More generally,
when one goes beyond an SU(6) quark model, there should be four independent
coupling constants, the $F$ and $D$ octet-octet ones, an octet-decuplet one
and a decuplet-decuplet one.

The results presented here are all obtained using a Gaussian form factor,
\begin{equation}
\mu(k)=\exp(-k^2/M^2),
\end{equation}
where the form-factor mass $M=660$ GeV has been fit to the pion-nucleon form
factor in the colour-dielectric model\cite{lb92}. We have also looked at other
form factors, for example the monopole form $1/(k^2+M^2)$, with form-factor
masses in the region of 1 GeV. These do not alter the qualitative behaviour of
our results.

We take the squares of the meson masses to be linearly related to the 
current quark masses (again ignoring isospin breaking):
\begin{eqnarray}
m_\pi^2&=&2B_0\overline m,\label{pimass}\\
m_K^2&=&B_0(m_s+\overline m),\label{kmass}\\
m_\eta^2&=&B_0\frac{2}{3}(2m_s+\overline m),\label{etamass}
\end{eqnarray}
where $\overline m$ is the average of the up- and down-quark current masses.
The chiral expansion of the self-energy in terms of the current masses is then
equivalent to one in terms of the meson masses.

The self-energy diagrams contain nonanalytic dependences on the meson
masses $m_\nu$\cite{gl82,gl84} and so the chiral expansion is not a
straightforward power series. These nonanalytic pieces correspond to infrared
divergences for some of the derivatives of the self energy in the chiral limit
where the mesons are massless. To illustrate how such an expansion can be
made, we consider first a simplified version of Eq.~(\ref{fullself}) in which
the bare baryon splittings are ignored. This corresponds to the part of the
self-energy of a nucleon arising from virtual $\pi$N states. The denominator
in the integrand is then just $k^2+m_\nu^2$.

To extract its nonanalytic parts, we first break the integral at some
arbitrary momentum $\Lambda$ into high- and low-momentum regions. We then take
the first $N$ terms of the expansion of the squared form factor in powers of
$k^2$,
\begin{equation}
\mu(k)^2 = \sum_{n=0}^\infty d_n k^{2n},
\end{equation}
and integrate these analytically (in practice using Mathematica\cite{math}).
This piece,
\begin{equation}
\Sigma_{AB\nu}^{(1)} =    
\frac{3}{4\pi^2}\left(\frac{g_{\pi NN}}{2M_N}\right)^2\frac{(f^{AB})^2}{25}
\sum_{n=0}^N d_n \int_0^\Lambda  \frac{k^{4+2n} dk}{k^2+m^2_{\nu}},
\label{sig1}
\end{equation}
contains all nonanalytic dependence on $m_\nu$ up to order $m_\nu^{2(N+1)}$.
The first such term is of order $m_\nu^3$, with a coefficient that can be 
checked against the standard ChPT result\cite{gl82}. The first logarithmic
dependence on $m_\nu$ appears at order $m_\nu^4\ln m_\nu$. 

The remaining low-momentum contribution,
\begin{equation}
\Sigma_{AB\nu}^{(2)} =  
\frac{3}{4\pi^2}\left(\frac{g_{\pi NN}}{2M_N}\right)^2\frac{(f^{AB})^2}{25}
\int_0^\Lambda\frac{k^4(\mu^2(k) - \sum_{n=0}^N d_n
k^{2n})dk}{k^2+m_{\nu}^2},
\end{equation}
can be expanded as a power series in $m_\nu$ to order $m_\nu^{2(N+1)}$ without
problems. The coefficients in this series involve integrals that must be
evaluated numerically.

Finally the high-momentum part of the integral 
\begin{equation}
\Sigma_{AB\nu}^{(3)} = 
\frac{3}{4\pi^2}\left(\frac{g_{\pi NN}}{2M_N}\right)^2\frac{(f^{AB})^2}{25}
\int_\Lambda^\infty \frac{k^4 \mu^2(k) dk}{k^2+m^2_{\nu}}
\label{sig3}
\end{equation}
can be safely expanded as a power series in $m_\nu$ since it is free of 
any infrared divergences. Again the coefficients are evaluated numerically.
When these three parts of $\Sigma_{AB\nu}$ are combined we find that, for 
our numerical integration, the sum is independent of $\Lambda$ over the range
200 to 800 MeV. This provides a useful check on both our analytic and
numerical calculations.

In Fig.~\ref{se8} we show the self-energy in this case as a function of 
the meson mass, along with the results of truncating the expansion at
orders up $m_\nu^7$. The convergence is very rapid for masses in the region of
$m_\pi$, with terms up to order $m_\nu^4$ accurately reproducing the full
integral. However for masses around $m_K$ the convergence is much slower, with
terms up to at least order $m^7$ being needed. The alternating signs of the
terms in the series are similar to what is seen in SU(3) ChPT calculations up
$m_\nu^4$\cite{bm96}. Those calculations show little evidence of convergence
up to that order.

The same techniques can be applied, with a little more effort, to the integral
with the full denominator in Eq.~(\ref{fullself}). In making the chiral
expansion we use the linear relations between the meson masses and the current
masses Eqs.~(\ref{pimass}--\ref{etamass}) and take the extra energy of a
strange quark, $\epsilon_s$ in Eq.~(\ref{mbare}), to be of order $m_s$, and
write it in the form 
\begin{equation}
\epsilon_s=a m_K^2.
\end{equation}

We also include the baryon decuplet in our calculations, with bare energy
splitting $\delta$. This splitting should remain finite in the chiral limit and
so is formally of chiral order $m^0$. Nonetheless it is numerically comparable
to the pion and kaon masses in size. An expansion in powers of those masses
for fixed $\delta$ would be very poorly convergent. We have therefore chosen to
make a simultaneous expansion of the self-energy in terms of the meson masses
$m_\nu$ and the splitting $\delta$, keeping terms to all orders in 
$m_\nu/\delta$\cite{jlms,jrt94,bm94}. In order to do this, we treat the bare
octet-decuplet splitting as if it were proportional to the pion or kaon mass,
writing it as
\begin{equation}
\delta=b m_\nu.
\end{equation}

The expansion of the full integral is made as described above, but with the
substitution
\begin{equation}
\frac{1}{(k^2+m_\nu^2)}\rightarrow \frac{1}{\omega(\omega+am_\nu^2+bm_\nu)}
\end{equation}
in the integrals of Eqs.~(\ref{sig1}--\ref{sig3}).
Some further care is needed if the splitting is larger than the meson mass,
$|b|>1$, which is the case for pion. Then the contribution to the self-energy
of the decuplet baryon arising from a pion-octet-baryon intermediate state has 
an imaginary part, reflecting the instability of that particle. The 
corresponding pion-decuplet-baryon contribution to the self-energy of the
octet baryon is of course purely real. Nonetheless its functional form differs
from that for smaller splittings. This can be seen most easily by considering
the expressions in the case $a=0$. Then the integrals contain
\begin{equation}
\frac{1}{\omega(\omega+bm_\nu)}=\frac{\omega+bm_\nu}{(k^2+(1-b^2)m_\nu^2)}.
\end{equation}
For $|b|<1$ the integrals corresponding to the two terms in the numerator of
this expression give rise to nonanalytic terms with a logarithmic dependence on
$m_\nu$. For $|b|>1$, there are no such logarithmic terms; instead the poles of
the integrands mean that both integrals develop imaginary parts. In the octet
case, $b$ is negative and these imaginary parts cancel exactly leaving a real
baryon mass. For any decuplet state except the $\Omega$, the imaginary parts
of the pion loop terms survive to leave a complex mass, reflecting the
unstable nature of these states.

The dependence of the self-energy on the meson mass in this case is shown in
Fig.~\ref{se10}. A bare octet-decuplet splitting of 300 MeV has been used. Also
shown are the results of a combined expansion in $m$ and $\delta$. For small
$m$ the dependence on $\delta$ controls the convergence of the expansion but
again keeping terms up to fourth order gives good accuracy in the region of
$m_\pi$. The convergence of the expansion is otherwise similar to that in
Fig.~1.

\section{Beyond the SU(6) quark model}

When baryon energies are evaluated as described in the previous section using 
the bare splittings and meson-baryon couplings of an SU(6) quark model, the 
results do not give a very good description of the observed 
spectrum\cite{mcg91}. For example, if the energy $\epsilon_s$ of a strange 
quark is chosen to reproduce the overall splitting of the baryon octet, then
the $\Sigma$-$\Lambda$ splitting is much too small. Alternatively, if the
$\Sigma$-$\Lambda$ splitting is reproduced, then the $\Xi$-$N$ is not. This is
illustrated by the first line of Table 2. We have therefore examined the
effect of more general coupling terms on our chiral expansion.

Such terms are routinely included in effective chiral Lagrangians for
meson-baryon physics (see, for example, \cite{jen92}). In a general chiral
Lagrangian, the leading-order couplings of mesons to octet baryons are of the
form
\begin{equation}
{\cal L}_{mBB} = 2 D Tr \bar{B} \gamma^{\mu} \gamma_5 \{A_{\mu},B\} +2 F Tr
\bar{B} \gamma^{\mu} \gamma_5 [A_{\mu},B],
\end{equation}
where $B$ denotes the octet baryon fields (expressed in $3\times 3$ matrix
form) and $A_{\mu}$ the axial current formed out the meson fields and their
derivatives. For a fuller definition of terms see\cite{jen92}. The
pion-nucleon coupling is related by the Goldberger-Treiman relation to the
axial coupling of the nucleon $g_A$, which in turn is given by
\begin{equation}
g_A = F + D.
\end{equation}
An SU(6) quark model would require a ratio $F/D=2/3$ but a somewhat smaller
value $F/D\simeq 0.57$-$0.58$\cite{fd} is deduced from data on semileptonic
decays of hyperons. We have therefore examined the dependence of our results
on $F/D$. As noted below, we find that small changes when $F/D$ is varied over 
a realistic range. We have therefore not considered the most general possible 
SU(3)-symmetric couplings which would involve independent octet-decuplet and
decuplet-decuplet coupling constants.

Of more importance is the replacement of the term $N_s\epsilon_s$ in 
the bare baryon masses of Eq.~({\ref{mbare}) by a more general
symmetry-breaking term. In the case of the baryon octet, two SU(3)-breaking
terms of octet form are possible:
\begin{equation}
{\cal L}_{\chi SB} = b_F Tr \bar{B} \left\{(\xi^\dagger M \xi^\dagger + \xi M
\xi),B\right\} +b_D Tr \bar{B} \left[(\xi^{\dagger} M \xi^{\dagger} + \xi M
\xi),B\right] 
\end{equation}
where $\xi$ is an SU(3) matrix constructed out of meson fields and $M$ is the 
meson mass matrix (again see\cite{jen92} for a complete definition). A
splitting proportional to strangeness, as in Eq.~(\ref{mbare}), is obtained if
only a $b_F$ term is included. The other term does not arise in simple SU(6) 
quark models, but can appear once gluon-exchange effects are included.
The strength of the splitting within the decuplet is also treated as an 
adjustable parameter, $b_{10}$.

In Table 2, we show the results for the baryon masses calculated using the 
full integrals, with $F/D=0.58$ and the other parameters chosen to fit the 
$\Xi$, average of $\Sigma$ and $\Lambda$, $\Delta$ and $\Xi^*$ masses. The 
corresponding values of the parameters are: $b_Fm_K^2=87.1$ MeV, 
$b_Dm_K^2=13.5$ MeV, $b_{10}m_K^2=129.0$ MeV and $\delta=318.0$ MeV. An overall
constant has been added to bring the nucleon mass up to its observed value.
A very good description of all eight masses is obtained (with five adjustable
parameters). The results are not sensitive to the $F/D$ ratio: if 2/3 is used
then the best fit value of $b_Dm_K^2$ is changed to 16.7 MeV and the other 
parameters are shifted by less than 1 MeV.

If one were to expand these results to first-order in the symmetry breaking
terms, one would get very different values for these masses. Nonetheless,
despite the importance of higher-order terms, the full masses continue to
satisfy relations that are often assumed to test the octet nature of the SU(3)
breaking terms in the baryon energies. For example, the GMO relation among the
octet baryon masses\cite{gmo} states that the combination
\begin{equation}
\Delta_{GMO}=\frac{3}{4}M_{\Lambda}+\frac{1}{4}M_{\Sigma}
-\frac{1}{2}M_N - \frac{1}{2}M_{\Xi}
\end{equation}
should vanish if the SU(3) breaking is purely octet in form. Similarly
for the decuplet one can construct two equal spacing rules (DES I and II
in the notation of\cite{djm94}):
\begin{equation}
\Delta_{DESI}=(M_{\Omega} - M_{\Xi^*}) - (M_{\Sigma^*} - M_{\Delta}),
\end{equation}
\begin{equation}
\Delta_{DESII}=\frac{1}{2}(M_{\Sigma^*} - M_{\Delta}) + \frac{1}{2}
(M_{\Omega} - M_{\Xi^*}) -M_{\Xi^*} + M_{\Sigma^*}.
\end{equation}
All three of these combinations of masses vanish exactly to leading order in 
the SU(3) symmetry terms, $b_F$, $b_D$ and $m_\nu^2$. 

For the baryon masses listed in Table 2, we find $\Delta_{GMO}=6$ MeV and
$\Delta_{DESI}=-2$ MeV, to be compared with the empirical violations of $6.6$
and $-13.6$ MeV respectively. The discrepancy between our result and the DES I
relation is largely due to the fact that our calculated $\Omega$ mass is out
by 10 MeV. The DES II relation, which has an empirical violation of $-3$ MeV,
is a rather poor test of the octet nature of the SU(3) breaking. Our one-loop
self energies would satisfy it exactly if we used the simple bare splittings of
Eq.~(\ref{mbare}) together with SU(3) symmetric meson-baryon couplings. As has
been noted by Jenkins\cite{jen92}, it is also satisfied exactly to order
$m_K^4$ in a chiral expansion. In a $1/N_c$ expansion violations of this
relation first appear at order $m_s^3/N_c^2$ and so are highly
suppressed\cite{jl95}. In our results the violation is very small, less than
0.1 MeV.

Thus despite the importance of higher-order terms which can transform under a
variety of representations of SU(3), the pattern of baryon masses remains
close to that expected from purely octet symmetry breaking, a point first made
by Jaffe\cite{jaf80} in the context of a chiral bag model. Hence the observed
success of the GMO relation cannot be used to infer that baryon masses can be
described using first-order perturbation theory in the current quark masses.
For example, to first order in the current quark masses the SU(3)-breaking
matrix element for the proton is given by
\begin{equation}
\frac{1}{3}(\overline m-m_s)\langle p|\bar uu+\bar dd-2\bar ss|p\rangle
\simeq M_\Lambda-M_\Xi=-202\ {\hbox{MeV}}.
\label{su3br}
\end{equation}
The contribution of the non-strange quark masses to the proton mass is given
by the $\pi N$ sigma commutator,
\begin{equation}
\sigma_{\pi N}=\overline m\langle p|\bar uu+\bar dd|p\rangle\simeq 45
\ {\hbox{MeV}},
\label{sigpin}
\end{equation}
where the value quoted is from the analysis of $\pi N$ scattering by Gasser 
{\it et al.}\cite{gls91}. Combining Eqs.~(\ref{su3br}) and (\ref{sigpin}) 
leads to an estimate of the contribution of the strange quark mass to the 
nucleon mass that is surprisingly large\cite{ch76}:
\begin{equation}
m_s\langle p|\bar ss|p\rangle\simeq\frac{1}{2}\left[{m_s\over \overline m}
\sigma_{\pi N}-3{M_\Xi-M_\Lambda\over 1-\overline m/m_s}\right]\simeq 260
\ {\hbox{MeV}}, 
\label{ssbar1}
\end{equation}
where the standard PCAC estimate of the ratio of quark masses\cite{gl82},
$m_s/\overline m\simeq 25$, has been used. Hence the use of first-order 
perturbation theory would suggest a large strange-quark content of nucleon.

Going beyond perturbation theory, the contributions of the quark masses to the
nucleon mass can be determined by applying the Feynman-Hellmann theorem. The
various contributions to the scalar quark densities from the meson clouds and
from intermediate-state baryons are:
\begin{eqnarray}
\overline{m} \langle N|\bar uu + \bar dd |N\rangle_{\pi} 
&=& m_{\pi}^2\frac{\partial\Sigma_N}{\partial m_{\pi}^2} \\ 
\overline{m} \langle N|\bar uu + \bar dd |N\rangle_{\eta} 
&=& \frac{m_\pi^2}{3} \frac{\partial\Sigma_N}{\partial m_{\eta}^2} \\ 
\overline{m} \langle N|\bar uu + \bar dd |N\rangle_K 
&=& \frac{m_\pi^2}{2} \frac{\partial\Sigma_N}{\partial m_K^2} \\ 
m_s \langle N|\bar ss|N\rangle_K &=& \frac{(2m_K^2 - m_{\pi}^2)}{2} 
\frac{\partial\Sigma_N}{\partial m_K^2} \\ 
m_s \langle N|\bar ss|N\rangle_\eta&=& \frac{3(m_\eta^2-m_\pi^2)}{4}
\frac{\partial\Sigma_N}{\partial m_{\eta}^2}\\ 
m_s \langle N|\bar ss|N\rangle_B &=& b_F\frac{\partial \Sigma_N}{\partial b_F}
+b_D\frac{\partial \Sigma_N}{\partial b_D}
+b_{10}\frac{\partial \Sigma_N}{\partial b_{10}}
\end{eqnarray}
The values of these are listed in Table 3.

We find similar results to Eq.~(\ref{ssbar1}) if we keep only terms of first
order in the symmetry-breaking. However the slow convergence of the expansion
in $m_K$ means that such large values for the strangeness content should not be
taken seriously. Indeed to next order ($m_K^3$) we find negative values for
$m_s\langle p|\bar ss|p\rangle$ (as in Ref.\cite{bkm93}) reflecting the
alternating signs of the terms in the self-energy that can be seen in Figs.~1
and 2. With the full loop integrals, the energy denominators ensure that kaon
and $\eta$ contributions to the self-energy are small, and hence that the
strange quark mass contributes only about 15 MeV to the nucleon mass. In
contrast the pion cloud contributes significantly to the $\pi N$ sigma
commutator, adding about 20 MeV to the $15-20$ MeV that the valence-quark
core would provide in a bag or soliton model. Including meson cloud effects
and treating SU(3)-breaking nonperturbatively, it is thus possible to have a
large $\sigma_{\pi N}$ without a large strangeness content in the nucleon, an
observation that has been made in the context of various bag and soliton
models\cite{jaf80,bmjtc,brsmb}.

\section{Conclusions}

We have examined the self-energies of the octet and decuplet baryons within
the framework of a generic cloudy-bag approach. These energies are calculated
at one-loop level from intermediate states consisting of a virtual pseudoscalar
meson and an octet or decuplet baryon. Meson-baryon form factors are used at
the vertices, and these regulate the loop integrals. In a chiral expansion of
the self-energies in powers of the meson masses, the form factors can be
thought of as providing model estimates for some of the higher-order
counterterms that would be present in a complete ChPT treatment.

The chiral expansion converges rapidly for the pion loops with intermediate
octet baryons, terms up to $m_\pi^4$ being sufficient to accurately reproduce
the full integral. For loops with intermediate decuplet baryons, the bare
octet-decuplet splitting $\delta$ provides a small energy denominator that
could lead to very poor convergence of a chiral expansion in $m_\pi$ alone.
Provided that one keeps terms to all orders in $m_\pi/\delta$ are kept, similar 
the contributions from pion-decuplet intermediate states show a similar
convergence to those from pion-octet states. In contrast the expansion of kaon
loops is much slower, with terms being needed up to order $m_K^7$ at least.
Our model for the higher-order counter terms of ChPT suggests that chiral
expansions are unlikey to be of much use for kaon-baryon physics, in contrast
to the situation for pions and baryons. This expectation is borne out by the
lack of convergence found in recent ChPT calculations to order 
$m_K^4$\cite{bm96}.

A good description of the octet and decuplet baryon spectrum can obtained in
this approach, provided that we include bare splittings that go beyond those
of a simple SU(6) quark model. Relations such GMO and the decuplet equal
spacing rules are very well satisfied by the calculated masses. Hence, despite
the need for high-order terms in the chiral expansion, the pattern of
splittings remains close to that produced by a purely octet SU(3)-breaking
term. The importance of such terms means that the observed $\pi N$ sigma
commutator cannot be used to deduce a large strangeness content of the nucleon.
In our treatment the pion cloud contributes nearly half of the observed sigma
commutator, yet the strangeness content of the proton is very small.

We believe that our approach provides reasonable estimates of the the 
higher-order terms in a chiral expansion. The results indicate that, while
such an expansion converges well enough to be useful for pion-nucleon physics,
this is not the case for kaons. Hence estimates of the strangeness content of
the proton based on low-order terms of a chiral expansion are not reliable.

\section*{Acknowledgments}
We are grateful to J. McGovern for many helpful discussions and for a critical
reading of the manuscript. This work is supported by the EPSRC and PPARC.


\begin{table}[h]
\begin{tabular}{l|l|l|l|l|l|l|l|l}
$A$ $B$ & $N$ & $\Sigma$ & $\Lambda$ & $\Xi$ & $\Delta$ & $\Sigma^*$ & $\Xi^*$ 
& $\Omega$ \\ 
\hline
$N$ & $9(F+D)^2$ & $\underline{9(F-D)^2}$ & $\underline{9F^2+6DF+D^2}$ & 
& 32 & $\underline{8}$ & & \\ 
& $9F^2-6DF+D^2$ & & & & & & & \\ 
\hline
$\Sigma$ & $\underline{6(F-D)^2}$ & $24F^2$ & $4D^2$ & $\underline{6(F+D)^2}$ & 
$\underline{\frac{64}{3}}$ & $\frac{16}{3}$ & $\underline{\frac{16}{3}}$ & \\ 
& & $4D^2$ & & & & 8 & & \\
\hline
$\Lambda$ & $18F^2+12DF+2D^2$ & $\underline{12D^2}$ & & $18F^2-12DF+2D^2$  
& & 24 & $\underline{16}$ & \\ 
& & & $\underline{4D^2}$ & & & & & \\
\hline
$\Xi$ &  & $9(F+D)^2$ & $9F^2-6DF+D^2$ & $\underline{9(F-D)^2}$ & 
& $\underline{8}$ & 8 & $\underline{16}$ \\ 
& & & & $\underline{9F^2+6DF+D^2}$ & & & 8 & \\
\hline
$\Delta$ & 8 & $\underline{8}$ & & & 25 & $\underline{10}$ & & \\ 
& & & & & 5 & & & \\
\hline
$\Sigma^*$ & $\frac{8}{3}$ & $\underline{\frac{8}{3}}$ & 4 & 
$\underline{\frac{8}{3}}$ & $\frac{40}{3}$ & $\frac{40}{3}$ & 
$\underline{\frac{40}{3}}$ & \\ 
& & $\underline{4}$ & & & & & & \\
\hline
$\Xi^*$ & & $\underline{4}$ & 4 & $\underline{4}$ & & 20 & 5 & 
$\underline{10}$ \\ 
& & & & $\underline{4}$ & & & $\underline{5}$ & \\
\hline
$\Omega$ & & & & $\underline{16}$ & & & 20 & \\ 
& & & & & & & & $\underline{20}$ \\
\end{tabular}
\smallskip
\smallskip
\caption{Squares of the SU(3) coefficients $f^{AB\nu}$ appearing in the
one-loop self energy of baryon $A$ involving an intermediate baryon state $B$
and a meson. In the strangeness-conserving entries the first line gives the
pion coefficient, the second the $\eta$ one. The underlined entries are those
for which $f_{AB\nu}$ is negative. In an SU(6) quark model the two octet-octet
couplings are given by $D=1$ and $F={2\over 3}$.}
\label{fab}
\end{table}

\begin{table}[h]
\begin{tabular}{l|llllllll}
& N & $\Lambda$ & $\Sigma$ & $\Xi$ & $\Delta$ & $\Sigma^*$ & $\Xi^*$ &
$\Omega$ \\ \hline
SU(6) & 939 & 1097 & 1193 & 1256 & 1232 & 1385 & 1535 & 1685 \\ 
Full  & 939 & 1115 & 1194 & 1318 & 1232 & 1384 & 1535 & 1684 \\ 
Expt. & 939 & 1116 & 1193 & 1318 & 1232 & 1385 & 1533 & 1672 \\ 
\end{tabular}
\smallskip
\smallskip
\caption{Baryon masses in MeV. The first line shows results of a calculation
using the bare mass splittings and meson-baryon couplings of an SU(6) quark
model. The second shows results of a calculation with $F/D=0.58$ and with the
general bare splittings of (3.3) adjusted to fit the observed
$\Sigma$-$\Lambda$ average, $\Xi$, $\Delta$ and $\Xi^*$ energies.}
\label{masses}
\end{table}

\begin{table}[h]
\begin{center}
\begin{tabular}{llllll}
$\overline{m}\langle u\bar{u}+d\bar{d}\rangle_{\pi} $ &
$\overline{m}\langle u\bar{u}+d\bar{d}\rangle_{K}$ & 
$\overline{m}\langle u\bar{u}+d\bar{d}\rangle_{\eta}$ & 
$m_s\langle s\bar{s}\rangle_K$ &
$m_s\langle s\bar{s}\rangle_\eta$ &
$m_s\langle s\bar{s}\rangle_B$ \\ \hline
22.7 & 0.7 & -0.02 & 16.7 & 0.5 & 6.5 \\ 
\end{tabular}
\smallskip
\smallskip
\caption{Contributions (in MeV) to the nonstrange and strange quark scalar
densities from the meson cloud and the intermediate baryons.}
\label{sigterms}
\end{center}
\end{table}

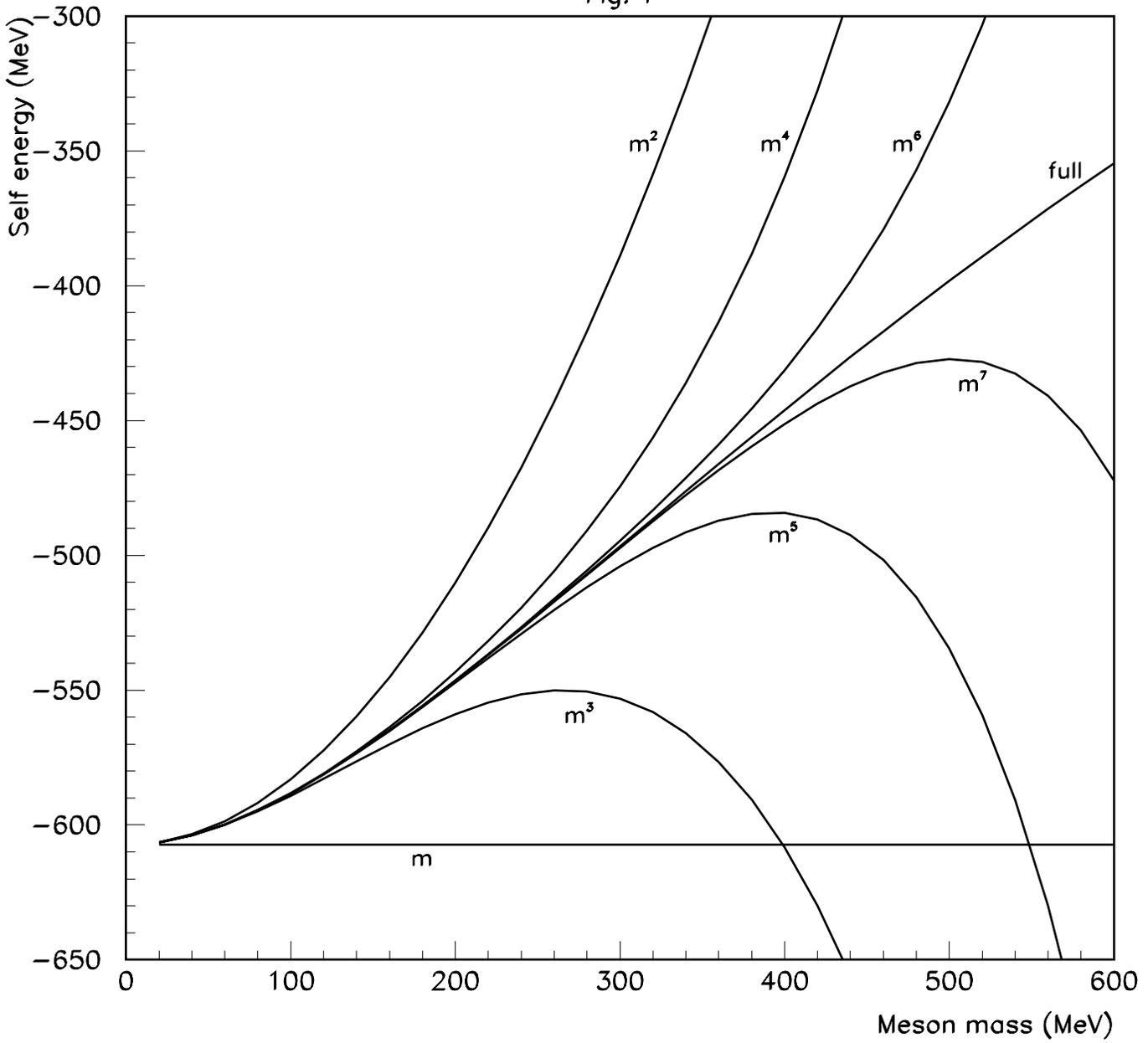
\begin{figure}
\caption{Dependence on meson mass of the octet baryon self-energy with
meson-octet-baryon intermediate states.}
\label{se8}
\end{figure}

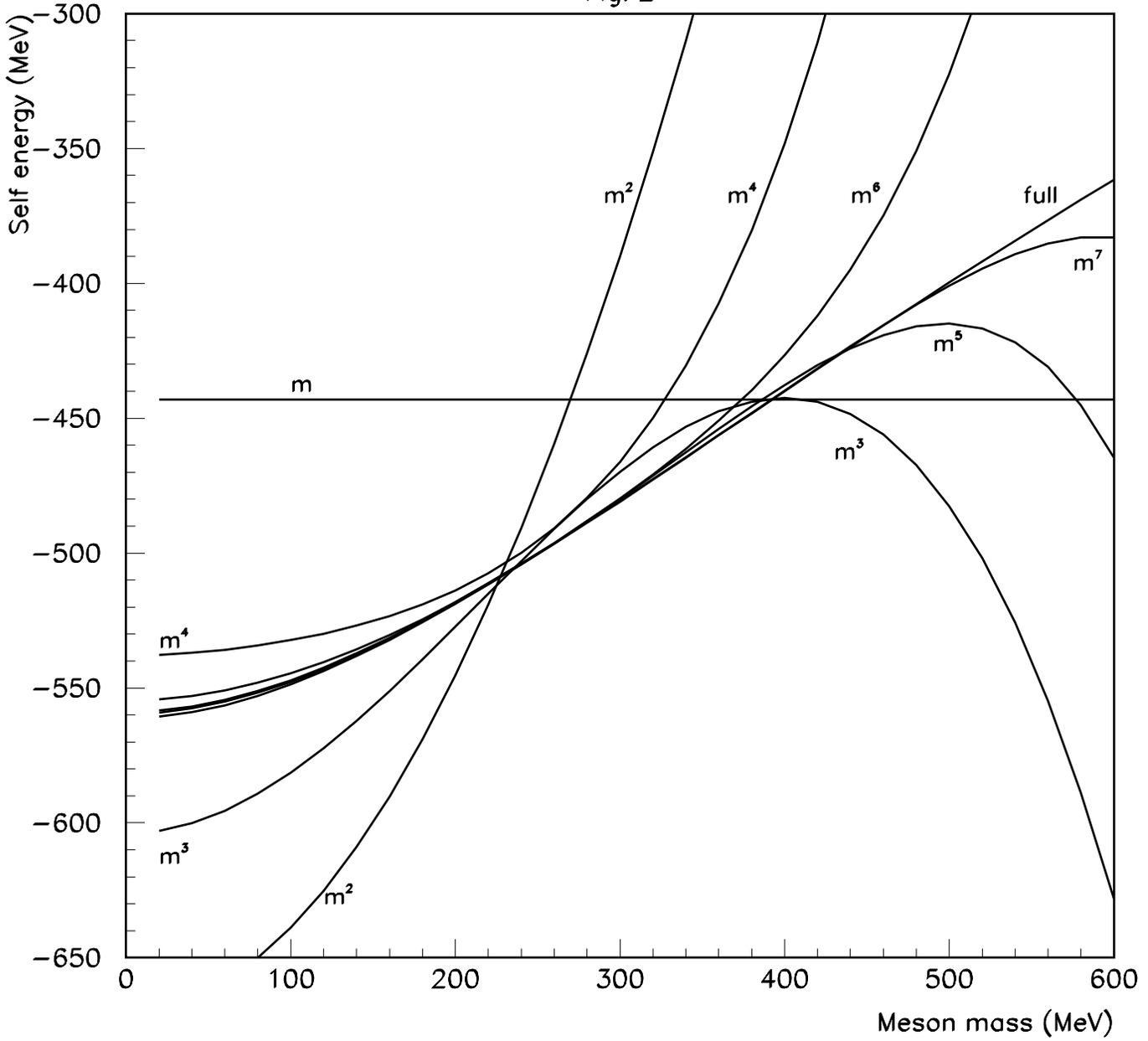
\begin{figure}
\caption{Dependence on meson mass of the octet baryon self-energy with
meson-decuplet-baryon intermediate states.}
\label{se10}
\end{figure}

\end{document}